\documentclass[a4paper,amsmath,amssymb,twocolumn]{revtex4}

\usepackage{bbm}

\def\be{\begin{equation}}
\def\ee{\end{equation}}
\def\ba{\begin{array}}
\def\bacc{\begin{array} {cc}}
\def\ea{\end{array}}
\def\bea{\begin{eqnarray}}
\def\eea{\end{eqnarray}}
\def\bd{\begin{displaymath}}
\def\ed{\end{displaymath}}

\begin{document}
\bibliographystyle{unsrt}

\title{Coupling Brane Fields to Bulk Supergravity}
\author{Susha L. Parameswaran}
\affiliation{Theoretical Physics,
  Uppsala University, Box 516, SE-751 20 Uppsala, Sweden, Tel: +46 (0)18 471 5955, Fax: +46 (0)18 53 31 80, Email:  susha.parameswaran@fysast.uu.se}
\author{Jonas Schmidt}
\affiliation{DESY Theory Group,
Notkestrasse 85, Bldg. 2a,
D-22603 Hamburg, Germany, Email:  schmijon@mail.desy.de} 

\begin{abstract}
In this note we present a simple, general prescription for coupling
brane localized fields to bulk supergravity.  We illustrate the
procedure by considering 6D N=2 bulk supergravity on a 2D orbifold,
with brane fields localized at the fixed points.  The resulting action
enjoys the full 6D N=2 symmetries in the bulk, and those of 4D N=1
supergravity at the brane positions. \\ \\

\noindent Keywords: supergravity, branes.
\end{abstract}

\maketitle

\section{Introduction}

Consider a bulk supergravity theory in higher dimensions ($D>4$), in
which the extra dimensions are compactified on an orbifold.  The
orbifold action has a number of fixed points, and certain fields may
be localized at these points.  In this way, we can construct a 4D
brane.  In the following we shall use the terms brane and fixed
point interchangeably.

At the fixed points, part of the higher dimensional gravitational- and
super-symmetries are broken explicitly.  For instance, half of the
supersymmetry generators are projected out.  But the subset of symmetries
corresponding to 4D N=1
supergravity survive.  Therefore, we can recast the bulk theory at the
brane in such a way that keeps 4D N=1 symmetries manifest.  4D N=1
supergravity is very well understood, and its general couplings to
matter was worked out in \cite{cremmer}.  So, we can use the
machinery of 4D N=1 supergravity to couple the bulk theory at the
brane to the localized fields.  The result will be a bulk plus brane
action, which enjoys the higher dimensional symmetries away from the
branes; the fields on the brane transform instead only under 4D N=1.  

We take our inspiration from \cite{sagnotti,mirabelli,arkani,stefan}
who reformulated higher dimensional globally supersymmetric theories
in terms of 4D N=1 superfields or their components.  Previous work on
brane-bulk couplings in local 5D models includes \cite{5D, ratazzi}.  One might have
believed that an off-shell description of the bulk supergravity be
necessary to construct general couplings to brane fields \footnote{For
recent work on the off-shell continuation of the local N=1 supergravity
into the bulk in the 6D case, see \cite{m10}.}.  This is
because on-shell, the supersymmetry algebra closes only up to the
equations of motion, and so care must be taken if we introduce new
terms to the action.  But we only add new interactions at the brane
positions where the N=2 supersymmetry is explicitly broken, and there we can invoke
by now text-book 
results of \cite{cremmer} on 4D N=1 supergravity.    

In detail, having recast the bulk theory at the brane into the form of
4D N=1 supergravity, we use the template of on-shell 4D N=1
supergravity with general matter couplings to write down the couplings
at the brane.  These general matter couplings were indeed first
derived via an off-shell formulation, but having established the
general on-shell Lagrangian there is no need to refer back to the
off-shell one.  Therefore, contrary to standard lore,  we are able to
simply use the component on-shell descriptions of both the higher
dimensional theory in the bulk and the 4D theory at the branes to
construct a general bulk-brane theory with the required symmetries.
On one hand, our method allows one to avoid the tedious (and possibly
never-ending) on-shell Noether procedure, and on the other hand we
can apply the method to cases where no off-shell description is
available, {\it e.g.} in 10D supergravity. 

We shall illustrate these ideas by coupling 4D N=1 brane fields to 6D
N=2 supergravity (with 8 supercharges).  That is, we also consider higher codimension branes.  One such model was constructed
by \cite{toffi,lee1,lee2}, who used the Noether method
to iteratively find appropriate brane-bulk couplings in the action and
supersymmetry transformation laws.  Our prescription has the
advantage that we can immediately introduce arbitrary 4D N=1 brane fields and their interactions, and moreover, it is also easy to write down the action up to four fermion terms.

Our motivation is to provide a field theory setup in which we can
study scenarios like Supersymmetric Large Extra Dimensions \cite{sled}, or
orbifold-GUTs \cite{orbGUTs}, allowing for non-trivial dynamics for
gravity, bulk and 
brane matter.    These constructions have provided interesting ways to
approach long-standing problems in cosmology and particle physics, and
find more fundamental descriptions within string 
theory.  For example, the brane fields may represent a field theory
description of the twisted sectors that arise in string orbifold
compactifications.  An intermediate 6D compactification \cite{jonas}
is particularly interesting in that context, since anisotropic
orbifolds allow an understanding of the mild hierarchy between the GUT
and Planck scales.   

Bearing in mind this purpose we shall make a few assumptions in order
to simplify our analysis and presentation.  Most of the work in our
construction goes in rearranging the bulk theory at the brane in terms
of 4D N=1 supergravity.  The constraint that odd fields, and internal
derivatives of even and odd fields, are vanishing at the brane
simplifies this task considerably.  In this way, we do not obtain all the
possible couplings between the bulk fields and our brane
fields.  We do however obtain the simplest ones that are necessary for
consistency with the symmetries, by which (charged) brane fields must
couple to the 4D metric (and gauge fields) and their N=1 supersymmetry
partners.  Moreover, we are able to immediately couple any possible
brane fields to the bulk.

\section{6D N=2 Bulk Supergravity}

We take a minimal on-shell field content in the bulk; a supergravity-tensor
multiplet $(e_M^A, \, B_{MN}, \, \varphi, \, \Psi^i_M, \, \chi^i)$, a
$U(1)$ vector multiplet $(A_M, \, \lambda^i)$ and
a charged bulk 
hypermultiplet $(\Phi^{\alpha}, \, \zeta^a)$.  We
take as the target quaternionic manifold of the hyperscalars the
canonical example 
$\frac{Sp(1,1)}{Sp(1) \times Sp(1)_R}$.  The action is (see Appendix A for our conventions) \cite{NS2}: 
\bea \label{SB}
    {\cal S}_B &=& \int d^6X e \left[-\frac{1}{2\kappa^2 } \, R + \frac{1}{2
      \kappa^2} \, 
    \partial_{M} \varphi \, \partial^M\varphi    \right. \cr
    && - \, \frac{1}{4} \, e^{\varphi}
    \; F_{MN}F^{MN} + \, \frac{1}{12}\, e^{2\varphi} \;
    G_{MNP}G^{MNP} \cr
    &&+ \frac{1}{2\kappa^2} \, g_{\alpha \beta }(\Phi) \,
    D_M \Phi^\alpha  \, D^M \Phi^\beta -  \, \frac{1}{2\kappa^4} \,
    e^{-\varphi} \, v(\Phi) \cr
&& \left. + {\rm fermions }\right] 
\eea
where the covariant derivative of the hyperscalars is:
\be
D_M \Phi^{\alpha} = \partial_M \Phi^{\alpha} - g A_M \xi^{\alpha}
\ee
with $\xi^{\alpha} = ({\rm T} \Phi)^{\alpha}$ the Killing vectors,
and ${\rm T}$ the Hermitian generator of the gauge group.  The
dependence of the potential on the hyperscalars
    is given by:
\be
v = P^{x} P^{x}
\ee
with $P^x$ the so-called Killing prepotentials, and $x$ running over
the adjoint of the composite
$Sp(1)_R$.  The prepotentials depend on the
spin-connection on the target manifold, ${\mathcal W}^x_{\alpha}$, and
the Killing vectors, as: 
\be P^{x}  =  g {\mathcal W}^x_{\alpha} \xi^{\alpha} 
\ee
and we give details on the target geometry in Appendix B.  The Kalb-Ramond field strength is given by $G_{MNP} =
    \partial_M B_{NP} + \frac{\kappa}{\sqrt{2}} F_{MN} A_P + 2 \,\,
	    {\rm perms}$.    

The fermionic supersymmetry transformations are (we shall always present
up to fermion bilinears only): 
\bea
&& \delta\Psi^i_M = \frac{\sqrt{2}}{\kappa} D_M\epsilon^i -
\frac{1}{24}e^{\varphi} 
  G_{NLR}\Gamma^{NLR}\Gamma_M\epsilon^i \cr
&& \delta\chi^i= -\frac{i}{\kappa\sqrt{2}}\partial_M\varphi
\Gamma^M\epsilon^i - 
  \frac{i}{12}  e^{\varphi} G_{MNL}\Gamma^{MNL}\epsilon^i \cr
&& \delta \lambda^{i} =  - \frac{1}{2\sqrt{2}}e^{\varphi/2}
 F_{MN}\Gamma^{MN} \epsilon^i -\frac{\sqrt{2}}{\kappa^2}e^{-\varphi/2}
        P^{x}({\rm T}^{x}\epsilon)^i \cr
&& \delta\zeta^a = \frac{i 
	  {\sqrt{2}}}{\kappa}(D_M\Phi^\alpha)\,V^{ai}_{\alpha}\Gamma^M\epsilon_i \,.    
\label{susy4}
\eea
Here,
$V^{ai}_{\alpha}$ is the vielbein on the target space manifold,
carrying the tangent space indices $a=1, 2$ and $i = 1,2$,
which run over the fundamental of the composite $Sp(1)$'s.  All spinors are
symplectic-Majorana Weyl, with the gravitino and gaugino being left-handed
and the dilatino and hyperinos being right-handed.  The
gravitini, Killing spinor, gaugini, and dilatini are all in the
fundamental of $Sp(1)_R$, whereas the gaugini and hyperini are charged under
the physical $U(1)$.  The
covariant derivative acting on  the Killing spinor is given by:
\be
D_M \epsilon^i = \partial_M \epsilon^i + \frac{1}{4} \omega_M^{AB}
\Gamma_{AB} \epsilon^i + (D_M
\Phi^{\alpha}) {\mathcal W}_{\alpha}^x {\rm T}^{xi}_j \epsilon^j.
\ee
 We
can always go to complex-Weyl spinors by defining $\epsilon =
\epsilon^1 + i \epsilon^2$ and so on.  We record the fermionic part of
the action and bosonic supersymmetry 
transformations in Appendix C.  Below we set $\kappa=1$.

Finally, let us note that as well as the gravitational- and
super-symmetries, our model has two kinds of gauge symmetries.  Under
the $U(1)$ gauge symmetry, not only do the gauge fields and
hypermultiplets transform, but also the Kalb-Ramond field due to
Cherns-Simons term in its Kalb-Ramond field strength (the latter must
clearly be gauge invariant):
\bea
&&A_M \rightarrow A_M + \partial_M \Lambda \cr
&&B_{MN} \rightarrow B_{MN} - \frac{\kappa}{\sqrt{2}} F_{MN} \Lambda \,.
\label{gaugexfmn}
\eea
Furthermore, there is an independent Kalb-Ramond gauge symmetry, whose
transformation is:
\be
B_{MN} \rightarrow B_{MN} + \partial_{[M} \Lambda_{N]} \, .\label{KRxfmn}
\ee 

\section{The Orbifold}
Let us now consider the bulk theory on a orbifold, $M/Z_2$.  $M$ is
a smooth, 2D manifold, for instance it could have topology
$\mathbb{R}^2$, or T$^2$ being a torus or a
deformed torus.  We can assign the
following parities with respect to 
the point group $Z_2$.  For the bosonic fields we choose: 
\bea
\text{even}&:& \,\, g_{\mu\nu}, g_{mn}, \varphi, \,\, B_{\mu\nu}, \,\, B_{mn}, \,\, A_{\mu}, \,\, \Phi^1, \,\, \Phi^3 \cr
\text{odd}&:& \,\,\, g_{\mu m}, \,\, B_{\mu m}, A_{m}, \,\, \Phi^2, \,\, \Phi^4,
\label{4dbosons}
\eea
and we can re-write the internal metric as:
\be 
    {g}_{mn} = \frac{r^2}{\tau_2}\begin{pmatrix}
    -1 & -\tau_1 \\ -\tau_1 &
    -\tau_1^2 - \tau_2^2 \end{pmatrix}. 
\ee
For the fermions, it is useful to first decompose the 6D complex Weyl spinors
into 4D ones as (see Appendix A for
  gamma matrix conventions):
\bea
    &&\Psi_M = \begin{pmatrix}
    \Psi_{L \,\mu} \\ \Psi_{R \, \mu}\end{pmatrix}, \,\,
  \begin{pmatrix} \Psi_{L
    \,m} \\ \Psi_{R \, m}\end{pmatrix}, \nonumber \\
&& \lambda = \begin{pmatrix}
    \lambda_L \\ \lambda_R\end{pmatrix}, \,\, \chi = \begin{pmatrix}
    \chi_R \\ \chi_L\end{pmatrix}, \,\, \zeta = \begin{pmatrix}\zeta_R \\
  \zeta_L\end{pmatrix},
\label{4dfermions}
\eea
and similarly for the 6D supersymmetry parameter, $\epsilon$:
\be
\epsilon = \begin{pmatrix}
    \epsilon_L \\ \epsilon_R\end{pmatrix}.
\ee
Then the corresponding fermionic parity assignments are:
\bea
\text{even:} \,\, && \Psi_{L \,\mu}, \,\, \Psi_{R \, m}, \,\, \chi_R, \,\,
\lambda_L, \,\, \zeta_R, \,\, \epsilon_L\cr 
\text{odd:} \,\,\, && \Psi_{R \,\mu}, \,\, \Psi_{L \, m}, \,\, \chi_L, \,\,
\lambda_R, \,\, \zeta_L, \,\, \epsilon_R. 
\eea
Notice that, although we have arranged the fields according to how
they transform under 4D general coordinate invariance, since we allow
them to depend on both external coordinates, $x^{\mu}$, and internal
coordinates, $y^m$, we are {\it not} dimensionally reducing.  Indeed,
the fields $r, \tau_1$ and $\tau_2$ carry all the degrees of freedom of
the extra dimensional components in the 6D metric. 

At the orbifold fixed points, a subset of the 6D N=2 symmetries are
explicitly broken.  In particular, the supersymmetry transformations
generated by the Killing spinor $\epsilon_R$ are projected out,
leaving only N=1 supersymmetry.  This is because the supersymmetry
parameter $\epsilon_R$ is a continuous odd function, and thus must be
vanishing at the fixed points.  Part of the 6D gravitational
symmetries are similarly broken, for example the general coordinate
transformations given by $x^m \rightarrow x^m + \xi^m$.  However, the
symmetries corresponding to N=1 4D supergravity survive, and some
additional ones.  These include part of the $U(1)$ gauge symmetries
(\ref{gaugexfmn}) and Kalb-Ramond gauge symmetries (\ref{KRxfmn}),
since $\partial_{\mu} \Lambda$ and $\partial_{[5}\Lambda_{6]}$ are
  non-vanishing on the brane. 

In the interests of simplicity we shall further assume that the odd
bulk fields, internal derivatives of even fields and internal
derivatives of odd fields are vanishing at the brane positions (unless
the symmetries require otherwise)
\footnote{One might wonder if the resulting dynamical problem is
  mathematically well-posed.  For example, in the 2D Cauchy Boundary
  Problem, both Dirichlet and Neumann boundary conditions are
  required.  Recall, however, that our branes have codimension two and
  do not represent boundaries in the internal dimensions, but points.
  We shall leave these formal issues aside.}.  
In the absence of brane sources, the first two of our conditions would
  be consequences of the orbifold parity symmetry, but couplings to
  brane sources may induce discontinuities, in the presence of which
  odd fields can be non-trivial at the fixed points \cite{bagger}.  With our
  assumptions, therefore, we will not obtain the most general
  couplings between bulk and brane fields.  We will, however, be able
  to couple general brane fields.  

One other comment on our constraints is in order, which is that they
also limit the possible background solutions that can be studied.
The constraint
that $\partial_m g_{\mu\nu}=0$ at the branes excludes some warp
factors, but those typically encountered in 6D brane world models
\cite{GGP, 8author, 100p} are allowed.        

\section{Bulk Theory at the Branes}
At the fixed points, the symmetries of 4D N=1 supergravity survive.  Therefore, the fields there assemble into on-shell N=1 supermultiplets.  For instance, the 4D scalars organize
into complex scalar components of N=1 chiral supermultiplets as $S =
\frac12 \left(s
+ i a \right)$, $T = \frac12 \left(t + i b\right)$, $U = \frac12 \left(\tau_2 + i 
\tau_1\right)$, $Z = \Phi^1 + i \Phi^3$ \cite{prekklt, toffi}, where we defined the scalars
$s=r^2 e^{\varphi}$ and $t= r^2 e^{-\varphi}$, and the psuedo-scalars
$a,b$ via:
\bea
&& G_{\mu\nu\rho} = \frac{r^{-4} e^{-2\varphi}}{\sqrt{2}} \,
\epsilon_{\mu\nu\rho\lambda} \partial^{\lambda}a, \cr
&& \partial_{\mu} b = \frac{1}{\sqrt{2}}\partial_{[\mu}
  B_{\dot{5}\dot{6}]} \,.
\eea
Here,  $\epsilon_{\mu\nu\rho\sigma}$ is the 4D Levi-Civita tensor.
Notice that we kept internal derivatives of the odd Kalb-Ramond field
components, since we must ensure invariance under the surviving parts of the
Kalb-Ramond gauge symmetry (\ref{KRxfmn}).  They do not however carry
independent physical degrees of freedom.     
The fermionic components of the chiral supermultiplets will be given
by three linear combinations of the fermions $\Psi_{R \, m}$ and 
$\chi_R$ as:
\bea
&&\psi^S =  \frac{{r^{3/2} \, e^{\varphi}}}{2} \, \left(\chi_R +
 \Psi_{R\,\dot{5}} -i \Psi_{R\,\dot{6}}\right) \cr
&&\psi^T =   \frac{{r^{3/2} \, e^{-\varphi}}}{2} \, \left( -\chi_R  +
 \Psi_{R\,\dot{5}} -i \Psi_{R\,\dot{6}}\right) \cr
&&\psi^U = \frac{r^{-1/2}{\tau_2}}{2} \, \left(\Psi_{R\,\dot{5}} +i \Psi_{R\,\dot{6}}\right)
\eea
and:
\be
\psi^Z = -\frac{\left(1-|Z|^2\right)}{2 r^{1/2}} \zeta_R \,.
\ee
Meanwhile, 
$A_{\mu}$ and $\lambda_L$ make up a N=1 vector multiplet. 
We will now
observe all this from the susy transformations.

Consider how the fermions at the branes transform under the N=1
supersymmetry that survives.  We write the corresponding
supersymmetry parameter as: 
\be
\epsilon= \begin{pmatrix} \epsilon_L(x) \\ 0 \end{pmatrix}.
\ee
It is a straightforward if laborious exercise to then rewrite the transformations
(\ref{susy4}) in terms of 4D 
fields defined above.  Remember that at the branes the odd fields and internal derivatives of odd and even fields vanish.  

At the same time, we Weyl rescale to the 4D Einstein frame, taking
$g_{\mu\nu} \rightarrow r^{-2} g_{\mu\nu}$.  Diagonalizing the
gravitino kinetic term, we find that the effective 4D gravitino on the brane is the
linear combination
$\psi_{L \, \mu} = \Psi_{L \, \mu} + \frac12 \Gamma_{\mu} \Gamma^m
\Psi_{R \,m}$.
Moreover, we perform the following chiral rotations on the fermions, in
order to obtain canonical kinetic terms; $\psi_{L \, \mu} \rightarrow
\psi_{L \, \mu}/r^{1/2}$ and $\lambda_L \rightarrow r^{3/2}
e^{\varphi/2} \lambda_L$, together with $\epsilon_L
\rightarrow 
\sqrt{2}\,\epsilon_L/r^{1/2}$.  We also find it convenient to scale out the
volume factor in the internal metric, $g_{mn} \rightarrow r^2 g_{mn}$.

After some beautiful cancellations, we find the following:
\bea
&& \delta\psi_{L \, \mu} = 2\,D_{\mu} \epsilon_L +
\frac{i}{2}\left(\frac{\partial_{\mu}a}{s} + \frac{D_{\mu}
    b}{t} +\frac{\partial_{\mu}\tau_1}{\tau_2} \right) \epsilon_L \cr
    && \qquad \quad + \frac{1}{1-|Z|^2} \left( Z D_{\mu} \bar{Z} - \bar{Z} D_{\mu} Z \right) \epsilon_L \cr
&& \delta \lambda_L = -\frac{1}{2} F_{\mu\nu} \gamma^{\mu\nu} \epsilon_L
 + i \frac{2\, g}{s} \frac{|Z|^2}{1-|Z|^2} \epsilon_L \cr
&& \delta \psi^S = -{i}\, \partial_{\mu}S \, \gamma^{\mu} \, \epsilon_L  \cr
&& \delta \psi^T = -{i}\, \partial_{\mu} T \, \gamma^{\mu} \, \epsilon_L \cr 
&& \delta \psi^U = -{i}\, \partial_{\mu} U \gamma^{\mu} \, \epsilon_L \cr
&& \delta \psi^Z = -i \, D_{\mu} Z \gamma^{\mu} \, \epsilon_L
\eea
where the complex scalar, $Z$, has charge +1 with respect to the
$U(1)$ gauge symmetry and so:
\be
D_{\mu} Z = \partial_{\mu} Z -i g A_{\mu} Z.
\ee
These 4D N=1 local supersymmetry transformations, along with the bulk Lagrangian evaluated at the brane positions, fall naturally within the general structure of 4D N=1 supergravity developed in \cite{cremmer}.
Indeed, at the branes, the bulk Lagrangian can be moulded into the form:
\bea
{\cal L}_B \vline_{b} = && e_4 \left[ -\frac12 R_{(4)} + K_{i \bar{j}} D_{\mu} \phi^i D^{\mu} \bar{\phi}^{\bar{j}} - \frac{1}{4} {\rm{Re}} H F_{\mu\nu} F^{\mu\nu} \right. \cr
&&  - \frac{1}{8} ({\rm{Re}} H)^{-1} \left( K_i {\rm T}^i_{\,j} \phi^j +
  h.c. \right)^2\cr
&& \left. + {\rm
    fermions} \right].  
\eea 
where $e_4$ and $R_{(4)}$ are the volume tensor density and Ricci scalar
associated with the 4D metric $g_{\mu\nu}$.  We have written the scalar
components of the N=1 chiral supermultiplets as $\phi^i=S, T, U, Z$,
and the K\"ahler potential is: 
\bea
K &=& -\log\left(T+\bar{T}\right) 
  -\log\left(S+\bar{S}\right) - \log\left(U+\bar{U}\right) \cr
&& -2 \log\left(1-Z\bar{Z}\right) \,.
\eea
Playing its role in the component Lagrangian, $K$ is a function of the scalar fields, and as usual, subscripts on $K$ indicate derivatives
with respect to the corresponding complex scalar.  The gauge kinetic
function, again a function of the scalar fields, can be identified as $H 
= 2S$.  Finally, the last term in the bosonic Lagrangian
takes the (on-shell) form of a D-term potential due to the charged
scalar.  The supersymmetry transformations 
similarly fall into the template of \cite{cremmer}: 
\bea
&& \delta\psi_{L \, \mu} = 2\,D_{\mu} \epsilon_L -
\frac{1}{2}\left(K_i D_{\mu} \phi^i - K_{\bar{i}} D_{\mu} \overline{\phi^i} \right) \epsilon_L \cr
&& \delta \lambda_L = -\frac{1}{2} F_{\mu\nu} \gamma^{\mu\nu} \epsilon_L
+ \frac{i}{2} ({\rm Re} H)^{-1} \left(K_i
{\rm T}^i_{\,j} \phi^j + h.c.\right) \epsilon_L \cr
&& \delta \psi^i = -i\, D_{\mu} \phi^i \, \gamma^{\mu} \, \epsilon_L \,.
\eea

\section{Bulk-Brane Couplings}
Having written the bulk theory at the brane in the standard form of
on-shell 4D N=1
supergravity, we can immediately couple any collection of on-shell 4D N=1 brane
fields localized at the fixed points, $y^m=y^m_b$.  This is because the general couplings in 4D N=1 supergravity have long been well understood, and these are the symmetries to be obeyed at the fixed points.  Indeed, the total Lagrangian at the fixed points, with contributions from the bulk and the brane fields, must take the form:
\bea
{\cal L}_B \vline_b + {\cal L}_b \, \delta^{(2)}(0) &=& e_4 \left[ -\frac12 R_{(4)} + K_{i \bar{j}} D_{\mu} \phi^i D^{\mu} \bar{\phi}^{\bar{j}} \right. \cr
&& - \frac{1}{4} {\rm{Re}} H_{(a)} F^{(a)}_{\mu\nu} F^{(a)\mu\nu} \cr
&& \left. - V_D - V_F + {\rm
    fermions} \right] \label{LBbgen}
\eea
where now $\phi^i$ and $A^{(a)}_{\mu}$ include any brane fields as
well as the bulk fields above, and we formally keep track of the
localization with delta-function distributions
$\delta^{(2)}(y^m-y^m_b)\equiv \delta(y^5-y^5_b) \delta(y^6-y^6_b)$,
where the superscript $^{(2)}$ indicates 
that we have 2D delta-functions.  In the Lagrangian,
there is a sum over 
gauge indices $(a)$, and the D-term potential is written as:
\be
V_D =  \frac{1}{8} ({\rm{Re}} H_{(a)})^{-1} \left( K_i {\rm T}^{(a)i}_{\,j}
\phi^j + h.c.\right)^2 \,.
\ee
We saw above that it may already have non-trivial contributions at
the classical level from the bulk.  There may be further contributions
to the scalar potential from the brane fields, to the D-term potential and to
an F-term potential, which we write in terms of the superpotential, $W$ (a function of the scalar fields), as:
\be
V_F = e^{K} \left(K^{i\bar{j}}D_i W D_{\bar{j}} \bar{W} - 3 |W|^2 \right) .
\ee
The supersymmetry transformations at the brane positions are similarly the standard ones of N=1 4D supergravity.  For the fermions, we have:
\bea
&& \delta\psi_{L \, \mu} = 2\,D_{\mu} \epsilon_L -
\frac{1}{2}\left(K_i D_{\mu} \phi^i - K_{\bar{i}} D_{\mu} \overline{\phi^i} \right) \epsilon_L \cr
&& \qquad \qquad \quad  - e^{K/2} W \gamma_{\mu} \overline{\epsilon}_L \cr
&& \delta \lambda^{(a)}_L = -\frac{1}{2} F^{(a)}_{\mu\nu} \gamma^{\mu\nu} \epsilon_L
  \cr
&& \qquad \qquad \,\,+ \frac{i}{2} ({\rm Re} H)^{-1} \left(K_i
{\rm T}^{(a)i}_{\,j} \phi^j + h.c.\right) \epsilon_L \cr
&& \delta \psi^i = -i \, D_{\mu} \phi^i \, \gamma^{\mu} \, \epsilon_L
- e^{K/2} K^{i\bar{j}} D_{\bar{j}} \overline{W} \overline\epsilon_L \,, \label{Bbsusygen}
\eea
and this clearly includes new terms, with respect to the original bulk transformations, depending on the brane fields.

Let us discuss two simple explicit examples, to illustrate the
generality of the scheme.  First, consider a brane localized chiral
supermultiplet, $(Q,\psi_Q)$, with 
charge +1 under the bulk $U(1)$ gauge symmetry and a canonical kinetic
term.  The total
Lagrangian at the brane positions takes the form (\ref{LBbgen}), with:
\bea
&& K = -\log\left(T+\bar{T}\right) -\log\left(S+\bar{S}\right) -
\log\left(U+\bar{U}\right) \cr 
&& \qquad \qquad \qquad -2\log\left(1-Z\bar{Z}\right) + Q\bar{Q} \delta^{(2)}(0) \cr
&& H = 2S \qquad \text{and} \qquad W=0 \,.
\eea 
Thus, the Lagrangian for the brane fields is given explicitly by:
\bea
{\cal L}_b &=& e_4 \left[g^{\mu\nu} D_{\mu} Q D_{\nu} \bar{Q} -
  \frac{1}{2s}g^2 |Q|^4 \delta^{(2)}(y^m-y_b^m)
  \right. \cr 
&& \left. \quad  - \frac{2g^2}{s} |Q|^2 \frac{|Z|^2}{1-|Z|^2} + {\rm fermions} \right] \, ,
\label{Lb1}
\eea
and we see that gauge invariance and local N=1 supersymmetry requires
the charged brane fields to couple not only to $g_{\mu\nu}$ and
$A_{\mu}$ but also to $s$ and
$Z$. It is also easy to observe from
(\ref{Bbsusygen}) that there are new brane localized field contributions
to the supersymmetry transformations of bulk fields, $\psi_{L \, \mu}$
and $\lambda_L$ (there are also new brane localized contributions to
$\delta\psi_{L \, \mu}$,$\delta\lambda_L$ and $\delta \psi^i$ at bilinear
order in the fermions, as can be read from the text-books). 
 
Take care that we have written the above couplings in terms of the
Weyl rescaled metric, corresponding to the 4D Einstein frame.  It is
in this frame that the bulk Lagrangian at the brane position takes the
standard 4D N=1 form.  If we want the couplings in terms of the
original 6D Einstein frame we must perform the inverse rescaling,
which leads to a further coupling between the brane fields and bulk
field $r$.  

The total Lagrangian is clearly invariant (up to total derivatives and the field equations) under the 6D N=2 local supersymmetries, with the brane localized fields transforming only under the 4D N=1 subset.  In detail, the Lagrangian and supersymmetry transformations are each composed of two parts; the original bulk supergravity interactions and the brane localized ones.    
The original supersymmetry variations of the bulk Lagrangian clearly cancel, since they are those of 6D N=2 supergravity.  The new brane localized contributions to the variation of the Lagrangian, which arise both due to new terms in the Lagrangian and the supersymmetry transformations, are by construction within the form of 4D N=1 supergravity with general matter couplings.  So they cancel too.  We have checked this explicitly for our simple example.
 Therefore, the total action: 
\be
{\cal S}_B + {\cal S}_b
\ee
with ${\cal S}_B$ given in (\ref{SB}) and 
\be
{\cal S}_b = \int d^6x {\cal L}_b \delta^{(2)}(y^m-y_b^m), 
\ee
is invariant
under the full 6D N=2 symmetries, with the brane fields $Q, \psi^Q$
transforming only under a 4D N=1 subset of them. 

 Moreover, it also follows easily that the supersymmetry algebra closes up to the field equations.   Observe again that the new brane localized terms that we have added in the supersymmetry transformations and field equations  correspond to standard interactions in 4D N=1 supergravity.   When computing the commutators of the supersymmetry transformations on the various fields,  we obtain purely bulk contributions and new brane localized contributions.  Applying the equations of motion introduces further bulk and brane localized contributions.  Finally, putting together all the terms, the bulk contributions close exactly as in 6D N=2 supergravity, and the brane localized contributions close exactly as in 4D N=1 supergravity.
 
Notice the presence
of a singular, delta-function squared term in the action, which is
required for the invariance of the action under the supersymmetry
transformations, and which is
typical in supersymmetric scenarios with localized fields
(see {\it e.g.} \cite{HW,mirabelli,arkani,5D,ratazzi,toffi,lee1,lee2,mirawarped}).  We shall discuss this a little more in our closing remarks.

As a second example, let us consider pure 6D N=2 supergravity in the
bulk, and introduce a  brane localized $U(1)$ gauge mutiplet, $(A_{\mu},\lambda)$, and a
charged brane 
chiral supermultiplet, $(Q,\psi_Q)$.  We allow the complex scalar $Q$ to have a kinetic coupling to the
bulk complex field, $T$.  This model will allow us to compare our
construction with the example worked out in the literature via the
Noether method \cite{toffi}.  It will prove convenient to redefine the
scalar component of the chiral supermultiplet, $T$, at the brane as: 
\be
T=t + |Q|^2 \delta^{(2)}(0)+ i b \, .
\ee
Notice that this implies a brane localized field contribution to the fermion $\psi^T$:
\bea
\psi^T =&&   \frac{{r^{3/2} \, e^{-\varphi}}}{2} \, \left( -\chi_R  +
 \Psi_{R\,\dot{5}} -i \Psi_{R\,\dot{6}}\right) \cr
&& + \left(\bar{Q} \psi^Q   + Q
 \psi^{\bar Q} \right)\delta^{(2)}(0) \, .
\eea 
The total Lagrangian at the brane can be chosen such that:
\bea
&& K = -\log\left(T+\bar{T} -2 |Q|^2  \delta^{(2)}(0)
\right) \cr
&& \qquad  -\log\left(S+\bar{S}\right) - \log\left(U+\bar{U}\right) \cr
&& H=1 \, \delta^{(2)}(0)  \qquad \text{and} \qquad W=0 \,.
\eea 
Then, the subsequent component Lagrangian for the brane fields is given explicitly by:
\bea
{\cal L}_b &=& e_4 \left[e^{\varphi} D_{\mu} Q D^{\mu} \bar{Q} -
  \frac{i}{2} e^{2\varphi} D_{\mu} b \left( Q D^{\mu} \bar{Q} -
  \bar{Q}D^{\mu} Q \right) \right.\cr 
&& - \frac{1}{4} e^{2\varphi} r^{-2} \left(Q D_{\mu} \bar{Q} - \bar{Q}
  D_{\mu} Q \right) \cr
&&\qquad \times \left(Q D^{\mu} \bar{Q} - \bar{Q} D^{\mu} Q \right) \delta^{(2)}(y^m-y_b^m)
  \cr 
&& \left. -\frac{1}{4} F_{\mu\nu} F^{\mu\nu} - \frac{1}{2}
  e^{-2\varphi} g^2 |Q|^4 + {\rm fermions} \right] \, , 
\eea
where $D_{\mu} Q = \partial_{\mu} Q + i g A_{\mu} Q$, and here we have
re-Weyl rescaled back to the 6D Einstein metric. 

The final result for the total action, ${\cal S}_B + {\cal S}_b$,
agrees with the one found in \cite{toffi}, \footnote{Indeed we can correct a
typo there in Equation (8), where the brane current should read
$j_{\mu} = \frac{i}{\sqrt{2}} \left(\bar{Q} D_{\mu} Q - D_{\mu}
\bar{Q} Q \right) + {\rm fermions}$).}.  Moreover, it is now easy to
complete the theory up to four fermion terms, and to show
that the bulk plus brane action does indeed have all the required symmetries:
6D N=2 in the bulk, 4D N=1 on the brane.

\section{Conclusions}
The idea that fields may be localized on a brane has played an
important role in many aspects of fundamental high energy physics and
cosmology for more than a decade.  However, building explicit,
detailed models which realize this idea remains technically
challenging, especially within the well-motivated framework of
supergravity.  
In the present letter, we have shown how to construct bulk plus brane
actions which incorporate the 
symmetries of 6D N=2 supergravity away from the branes and
4D N=1 supergravity at the brane positions.  The power of our
approach is that we do not need to enter into extremely lengthy and
messy computations to check the supersymmetry invariance and closure
of the algebra each time we add new brane localized fields and interactions.
Instead, one can simply  
write down N=1 preserving interactions between the bulk supergravity
and brane localized fields because the general matter couplings for 4D
N=1 supergravity are known, and the bulk theory at the brane positions can
be recast into that form. 

It is this latter step that represents the technical challenge in our
proposal, and we reduce it by making some simplifying assumptions for
the behaviour of bulk fields at the branes, including that the internal
derivatives are vanishing there.  We are still able to
couple general brane fields to the bulk theory, {\it e.g.} it becomes
very easy to extend the model of Ref \cite{toffi}, first constructed by the
Noether method.  However, it would certainly
be interesting to relax those assumptions, to allow the most general
bulk-brane couplings possible.  Of course, in principle, it should
indeed be possible and interesting to rewrite the whole 6D N=2 bulk
theory keeping only the N=1 supersymmetries manifest (see \cite{m10}
for some first steps). 

Our focus has been on the bosonic part of the action and its fermionic
supersymmetry transformations, since this is most interesting part.
The completion to the fermionic action and bosonic transformations is
of course guaranteed by supersymmetry, and can be read from the
text-books.  At the same time, our analysis 
has been entirely classical.  At the quantum level, there are
generically gauge anomalies in the bulk and on the brane, as well as
bulk gravitational anomalies, and these provide restrictions on the
bulk and brane matter contents.  

The results presented allow us to build field theory models
describing, for example, 
the low energy dynamics of orbifold string compactifications.  For
instance, our construction is rich enough to build supergravity
realizations of the orbifold-GUT models in \cite{orbGUTs}.  Orbifold
string compactifications have proved remarkably sucessful in the quest for a
fundamental origin of the MSSM \cite{mssm}, but their dynamical
aspects remain to 
be understood.  Ref \cite{kai1, kai2} suggest that a non-trivial
dynamics for the brane localized fields may help in stabilizing the
bulk moduli, using a toy globally supersymmetric 6D model.
We may now study such issues taking into proper account the
consequences of dynamical gravity.  

As was to be expected from previous work (starting with \cite{HW}), our supergravity actions
describing the brane localized fields suffer from the presence of
delta-function squared terms, which are indeed required by
supersymmetry.  It is generally believed that such 
singularities would be resolved in a full quantum gravity
treatment.  At the same time, we can pragmatically try to live with
them within the effective field theory \cite{HW} (see \cite{mirabelli}
for a useful representation of them).  For example, we
expect them to play an 
important role in the cancellation of divergences ensured by
supersymmetry \cite{mirabelli}.  This was shown explicitly for a
global 5D model in 
\cite{mirabelli,mirawarped}, and also for a 5D supergravity model in
\cite{ratazzi}.  It would furthermore be important to
develop techniques  
to construct non-trivial background compactifications and subsequent low energy
4D effective field theories, despite these
singularities.  For recent work on the backreaction of codimension-two
branes in the absence of brane matter see \cite{codim2}.  Finally, for
some insights regarding the subtleties of supersymmetry 
in singular spaces, we refer to \cite{kallosh}.

\begin{acknowledgements}
We would like to thank Wilfried Buchm\"uller, Hyun Min Lee, Jan Louis
and Jan Moeller for several discussions and comments on the
manuscript.  We also thank Christoph Ludeling and Stefan Groot
Nibbelink for discussions, and acknowledge Jan Moeller for
collaboration on related ideas.  Finally, we are grateful to the anonymous
referee for their comments.  S.L.P is supported by the G\"{o}ran
Gustafsson Foundation. 
\end{acknowledgements}

\appendix

\section{Conventions}
Our signature is mostly minus and we take MTW conventions for the
curvature tensors.  6D spacetime coordinates are $X^M$, 4D ones are
$x^{\mu}$ and 2D ones $y^m$.  Tangent space indices are, respectively,
$A, B, \dots$; $\alpha, \beta, \dots$ and $a, b, \dots = \dot{5},
\dot{6}$. 

We build the 4D gamma matrices from the Pauli matrices,
\be
\sigma^1 = \begin{pmatrix} 0 & 1 \\ 1 & 0 \end{pmatrix}, \quad \sigma^2 = \begin{pmatrix} 0 & -i \\ i & 0 \end{pmatrix} \quad \sigma^3 = \begin{pmatrix} 1 & 0 \\ 0 & -1 \end{pmatrix},
\ee
as follows:
\be
\gamma^0 = \begin{pmatrix} 0 & \mathbbm{1}_2 \\ \mathbbm{1}_2 & 0 \end{pmatrix}, \, \gamma^i = \begin{pmatrix} 0 & \sigma^i \\ -\sigma^i & 0 \end{pmatrix}, \, \gamma^5 = \begin{pmatrix} -\mathbbm{1}_2 & 0 \\ 0 & \mathbbm{1}_2 \end{pmatrix}.
\ee
In turn, the 6D gamma matrices are:
\be
\Gamma^{\alpha} = \begin{pmatrix} \gamma^{\alpha} & 0 \\ 0 & \gamma^{\alpha} \end{pmatrix}, \, \Gamma^{\dot{5}} = \begin{pmatrix} 0 & i \gamma_5 \\ i \gamma_5 & 0 \end{pmatrix}, \, \Gamma^{\dot{6}} = \begin{pmatrix} 0 & \gamma_5 \\ -\gamma_5 & 0 \end{pmatrix},
\ee
with the chirality matrix
\be
\Gamma_7 = \begin{pmatrix} \gamma_5 & 0 \\ 0 & -\gamma_5 \end{pmatrix}.
\ee
The 2D Levi-Civita tensor density is $\epsilon_{ab}$, with
$\epsilon_{\dot{5}\dot{6}}=1$.

\section{Hyper Target Geometry}
The hyperscalars in N=2 6D supergravity coordinatize a quaternionic
manifold, and we shall take as a canonical example the
manifold $\frac{Sp(1,1)}{Sp(1) \times Sp(1)_R}$.  The geometry of this
class of manifolds is described in detail in \cite{superswirl}.  With
our four hyperscalars, we can compose a quaternion, $t=\Phi^1 {\bf 1}
+ \Phi^2 {\bf i} + \Phi^3 {\bf j} + \Phi^4 {\bf k}$, where we have
introduced the following $2 \times 2$ basis for the quaternions: 
\bea \label{E:qbasis} {\bf i} &=&  \left(
     \begin{array}{cc}
      -i & \,\,0 \\
      \,\,0 & \,\,i  \end{array}
     \right) , \qquad \qquad
{\bf j} = \left(
     \begin{array}{cc}
      0 & -1 \\
      1 & \,\,0 \end{array}
     \right) , \cr 
{\bf k} &=& \left(
     \begin{array}{cc}
      \,\,0 & -i \\
      -i & \,\, 0 \end{array}
     \right) ,\qquad \qquad
{\bf 1} = \left(
     \begin{array}{cc}
      \,\, 1 & \,\,0 \\
      \,\, 0 & \,\,1 \end{array}
     \right).
\eea
The $Sp(1)_R$ spin-connection and vielbein are given in terms of the
      quaternion by:
\bea \label{E:targetgeom}
{\mathcal W}_{\alpha}^{ij} &=& \frac{1}{2} \gamma^{-2} \left(
\partial_{\alpha}
      t^{\dagger} \, t - t^{\dagger} \partial_{\alpha}t \right) \cr
V_{\alpha}^{ai} &=& \gamma^{-1} \left( I - tt^{\dagger}
\right)^{-1/2}
      \partial_{\alpha}t .
\eea
The target manifold metric is then given explicitly by:
\be
g_{\alpha\beta}= \frac{2}{1-|\Phi|^2} \delta_{\alpha\beta} ,
\ee
with the shorthand $|\Phi|^2 = (\Phi^1)^2+(\Phi^2)^2+(\Phi^3)^2+(\Phi^4)^2$.
Meanwhile, we choose the hypermultiplet to be charged under the bulk
      $U(1)$ gauge symmetry, such that the Killing vector
      $\xi^{\alpha}=({\rm T}\Phi)^{\alpha}$ is:
\be
\xi^{\alpha}= \left( \begin{array}{c} -\Phi^3 \\ \Phi^4 \\ \Phi^1 \\ -\Phi^2 \end{array} \right).
\ee
The Killing prepotentials are then given by:
\be
P^x {\rm T}^{x \, ij} \equiv P^{ij}= g \frac{|\Phi|^2}{1-|\Phi|^2} \left( \begin{array}{cc}
      \,\, 0 & \,\,1 \\
      \,\, -1 & \,\,0 \end{array}
     \right).
\ee
Using $Tr {\rm T}^x {\rm T}^y = - \frac12 \delta^{xy}$, we find that the 6D scalar
      potential function is:
\be
v(\Phi) = 4 g^2 \frac{|\Phi|^4}{(1-|\Phi|^2)^2}
\ee

\section{Fermionic Action and Bosonic SUSY Transformations}
The fermionic part of the action for minimal 6D supergravity, to
bilinear order, is:
\bea
S_F &=&\int d^6X e_6 \left[ -\frac12 i\, \bar{\Psi_M} \Gamma^{MNP} D_N
  \Psi_P + \frac12 i\, \bar{\chi} \Gamma^M D_M \chi \right.\cr
&& + \frac12 i\,
  \bar{\lambda} \Gamma^M D_M \lambda + \frac12 i \bar{\zeta^a} \Gamma^M
  D_M \zeta_a  \cr
&&\frac{1}{4} \bar{\chi} \Gamma^N \Gamma^M \Psi_N
  \partial_M \varphi + \frac12 \bar{\Psi_M^i}\Gamma^N\Gamma^M \zeta^a
  V_{\alpha a i} D_N \Phi^{\alpha} \cr
&& -\frac{\kappa}{48\sqrt{2}} i\, e^{\varphi} G_{MNP}
\left( - \bar{\chi}
  \Gamma^{MNP} \chi + \lambda \Gamma^{MNP} \lambda \right. \cr
&& \qquad + \bar{\zeta^a} 
  \Gamma^{MNP} \zeta_a \cr
&& \left. \qquad -\bar{\Psi^L}\Gamma_{[L}\Gamma^{MNP} \Gamma_{S]} \Psi^S +
  2\,i\,\bar{\Psi_L} \Gamma^{MNP} \Gamma^{L} \chi \right) \cr
&& + \frac{\kappa}{8}\ i\, e^{\varphi/2} F_{PQ} \left({\bar
    \Psi_M} \Gamma^{PQ} \Gamma^M \lambda - i \chi \Gamma^{PQ} \lambda
\right) \cr
&&  e^{-\varphi/2} i \kappa^{-1} \sqrt{\frac12} \left(\bar{\Psi}_M \Gamma^M
    T^x \lambda C^x \right. \cr
&& \left. \left.\qquad \qquad \quad + i \bar{\chi} T^x \lambda C^x + 2\,i\,\zeta^a
    \lambda^i V_{\alpha a i} \xi^{\alpha} \right) \right] \,.
\eea
Meanwhile, the supersymmetry transformations for the bosonic fields,
up to fermion bilinears, are:
\bea
&& \delta e_M^A = -i \, \frac{\kappa}{\sqrt{2}} \bar{\epsilon}
\Gamma^A \Psi_M \cr
&& \delta \varphi = \frac{\kappa}{\sqrt{2}} \bar{\epsilon} \chi \cr
&& \delta B_{MN} = -\frac12 i\, e^{-\varphi}\left( \bar{\epsilon}
  \Gamma_M \Psi_N - \bar{\epsilon} \Gamma_N \Psi_N - i\,
  \bar{\epsilon} \Gamma_{MN} \chi \right) \cr
&& \qquad \qquad + \sqrt{2} \kappa A_{[M} \delta A_{N]} \cr
&& \delta A_M = -\sqrt{\frac12} i\, e^{-\varphi/2}
\bar{\epsilon}\Gamma_M \lambda \cr
&& \delta \Phi^{\alpha} = -\frac{\kappa}{\sqrt{2}}
V^{\alpha}_{ai}\bar{\epsilon^i} \zeta^a \,.
\eea

\section{Metric Decomposition}
Here, we write down the decomposition of the bulk curvature
tensors and connections, which appear in the intermediate steps of our
calculations.  Recalling the Weyl rescalings, we decompose the 6D
metric as:
\be
    {g}_{MN} = \begin{pmatrix}
    r^{-2}{g}_{\mu\nu} & 0 \\ 0 &
    r^{2}{g}_{mn} \end{pmatrix},
\ee
with the corresponding vierbiens, $e_{\mu}^{\alpha}$ and
\be
    e_m^{\,\,a} = \begin{pmatrix}
    -\frac{1}{\sqrt{\tau_2}} & 0 \\  -\frac{\tau_1}{\sqrt{\tau_2}} &
    \sqrt{\tau_2} \end{pmatrix}.
\ee 
The relevant components of the 6D spin-connection, $\omega_M^{AB}$, are then:
\bea
&& \omega_{\mu}^{\alpha\beta} =
\omega_{\mu}^{\alpha\beta}[e_{\mu}^{\alpha}] + e^{\nu\beta}
e_{\mu}^{\,\,\alpha} \partial_{\nu} r^{-1} -  e^{\nu\alpha}
e_{\mu}^{\,\,\beta} \partial_{\nu} r^{-1}\cr
&& \omega_{\mu}^{\alpha b} = \text{internal derivs of} \,\, r \cr
&& \omega_{\mu}^{\dot{5}\dot{6}} = -  \omega_{\mu}^{\dot{6}\dot{5}}=   \frac{\partial_{\mu}\tau_1}{2  \tau_2} \cr
&& \omega_{m}^{\alpha\beta} = 0\cr
&& \omega_{m}^{\alpha a} = r e^{\mu\alpha}\begin{pmatrix} -\frac{\partial_{\mu}
    \tau_2}{2 \tau_2^{3/2}} & -\frac{\partial_{\mu}\tau_1}{2 \tau_2^{3/2}} \\  \frac{\tau_2 \partial_{\mu} \tau_1 -
    \tau_1 \partial_{\mu} \tau_2}{2 \tau_2^{3/2}} &  -\frac{\tau_1 \partial_{\mu} \tau_1 +
    \tau_2 \partial_{\mu} \tau_2}{2 \tau_2^{3/2}} \end{pmatrix}
 \cr 
&& \qquad \qquad +
e^{\mu\alpha} e_m^{\,\,a} r \partial_{\mu}r \cr
&&  \omega_{5}^{\dot{5}\dot{6}} = - \omega_5^{\dot{6}{\dot{5}}} = 
\text{internal derivs of} \,\, r,
\, \tau_1, \, \tau_2 \cr
&& \omega_6^{\dot{5}\dot{6}} = - \omega_6^{\dot{6}\dot{5}} = 
\text{internal derivs of} \,\, r,
\, \tau_1, \, \tau_2 \, .
\eea
Subsequently, the 6D Ricci scalar decomposes as:
\bea
R &=& r^2 R_{(4)}[g_{\mu\nu}] - \frac{1}{2 \tau_2^2}
g^{\mu\nu} \partial_{\mu}\tau_2 \partial_{\nu} \tau_2 -  \frac{1}{2 \tau_2^2}
g^{\mu\nu} \partial_{\mu}\tau_1 \partial_{\nu} \tau_1 \cr
&&  - \frac{1}{r^4} g^{\mu\nu}
\partial_{\mu} r^2 \partial_{\nu} r^2 \cr
&& + \text{internal derivs of} \,\, r,
\, \tau_1, \, \tau_2 \, .
\eea

\end{document}